\documentclass[aps,pra,twocolumn,footinbib,floatfix]{revtex4}
\usepackage{amssymb}
\usepackage[pdftex]{graphicx}	
	\DeclareGraphicsExtensions{.pdf,.png,.jpg}
\usepackage[applemac]{inputenc}
\usepackage{amsmath,amssymb,amsthm}
\usepackage{mathbbol,bbm}
\usepackage{color}
\usepackage{graphicx,float}
\usepackage[final]{showkeys}
\usepackage{bm,dsfont}
\usepackage{tensor}

\def\ket#1{\left|#1\right>}
\def\bra#1{\left<#1\right|}

\usepackage{booktabs}

\begin{document}
\title{On the role of joint probability distributions of incompatible observables in Bell and Kochen-Specker Theorems}
\author{\'Angel Rivas}

\affiliation{Departamento de F\'isica Te\'orica, Facultad de Ciencias F\'isicas,
Universidad Complutense, 28040 Madrid, Spain.\\
CCS -Center for Computational Simulation, Campus de Montegancedo UPM, 28660 Boadilla del Monte, Madrid, Spain.}

\vspace{-3.5cm}

\begin{abstract}
We analyze the validity of Bell and Kochen-Specker theorems under local (or noncontextual) realism but avoiding an assumption of the existence of a joint probability distribution for incompatible observables. We formulate a realist model which complies with this requirement. This is obtained by employing divergent sequences that nevertheless have marginals which are convergent. We find that under standard reasonable assumptions this possibility does not lead to a loophole of those theorems, by deriving a short of CHSH inequality valid for any finite size ensemble. Moreover, we analyze a Hardy's paradox setting where noncontextual realism imposes the existence of joint probabilities for incompatible observables.
\end{abstract}


\maketitle


\section{Introduction}
It is common knowledge that quantum mechanics does not allow to predict the result of individual measurements but only the probabilities associated to each of the possible results. The origin of this randomness has been subject to much debate since the early days of the quantum theory \cite{EPR}. While there is as yet no definitive resolution, there are some strong results which shows that if a more fundamental theory explaining the statistical nature of the quantum physics
exists, some well motivated physical principles must be contravened. For example, because of the violations of Bell inequalities \cite{Bell}, it is argued that it not possible to find a theory ``completing'' quantum mechanics keeping the causal structure of the Einstein's relativity (nonlocality). Another result originally due to Kocher and Specker (KS) \cite{KS} claims that in such a more fundamental theory the value of some observable cannot depend only on the state of the system, but also on the particular experimental setting used to measure it (contextuality). For convenience, we will often refer to tests of Bell and/or of Kochen-Specker theorems as Bell-KS tests.

However, the arguments behind these no-go theorems are often presented in a way where these assumptions interweave with other well-established (and tacitly assumed) beliefs. This has been and is a cause of debates regarding the ultimate reason why quantum mechanics violates these no-go results. One of these beliefs is the potential existence of a joint probability distribution for incompatible observables \cite{Fine1,Fine2}.  

On the one hand, it is well-known that quantum mechanics does not violates Bell-KS tests if there were a well-defined and positive probability joint distribution of all observables involved in these tests. This is guaranteed for commuting observables, so that they are compatible. Thus, in order to obtain a contradiction some incompatibility (i.e. noncommutability) is needed. As a result, the violation of Bell-KS tests could be attributed to the absence of well-behaved joint probability distributions, in combination with assumptions such as realism, locality/noncontextuality and free will. The purpose of this work is to address this question in detail and explain how it is possible to derive these theorems just from the assumptions of realism, locality/noncontextuality and free will, even for models with ill-defined joint probability distributions for noncommuting quantum observables. As we shall see this does not fall into mathematical contradiction with the Fine's results \cite{Fine1,Fine2}. 

The paper is organized as follows. The next two sections review the basic assumptions involved in Bell-KS theorems, and the relation between hidden variable models with joint distributions, respectively. In Sec. IV we formulate a realist model which lacks joint distributions for incompatible observables. Then in Sec. V we analyze whether such a realist model can be used to describe the result of quantum experiments. For that purpose we derive a short of CHSH inequality \cite{CHSHpaper} valid for physical ensembles with a finite number of systems. The extension of Bell inequalities for finite size sampling has appeared in the literature in different contexts, e.~g. \cite{Stapp,Mermin2,Macdonald,Gill}, we discuss the connection with these works in the final discussion section. Moreover, in Sec. VI we present an argument based on the Hardy paradox \cite{Hardy1} which forces the existence of joint probability distributions for incompatible observables for a particular quantum state.  The work is completed with an Appendix which includes some mathematical detailed derivations used in the main sections.

\section{Realism, determinism, locality, noncontextuality, and measurement independence}

\begin{figure}[t]
\begin{center}
\includegraphics[width=1\columnwidth]{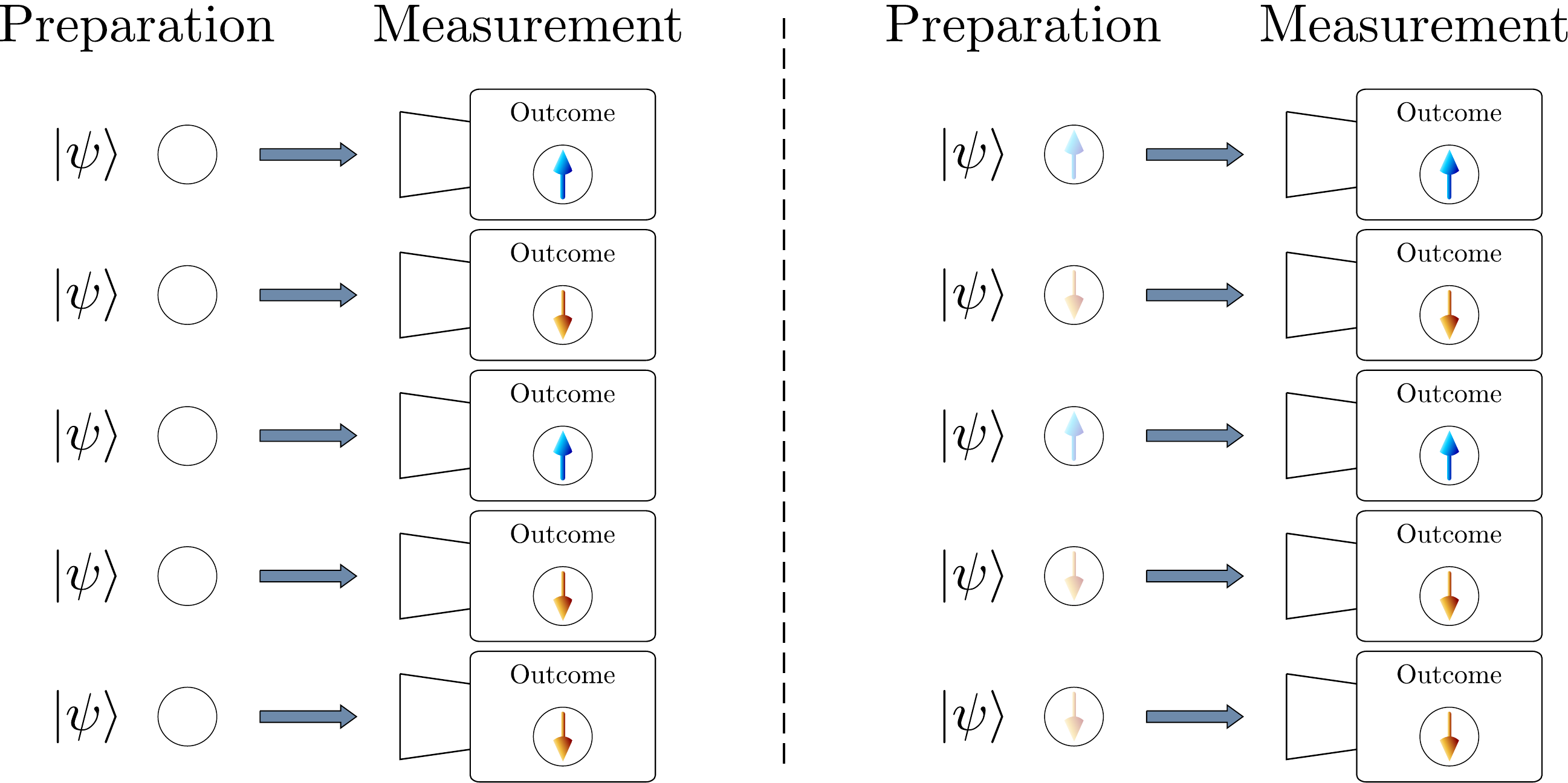} 
\end{center}
\caption{Illustration of a realistic view of a quantum experiment. On the left left hand side: a 1/2-spin system is prepared in some quantum state $|\psi\rangle$ and a measurement of some spin component, say $s_z$, is performed obtaining ``$\uparrow$'' or ``$\downarrow$''. The experiment is repeated under identical conditions obtaining the sequence of outcomes ``$\uparrow,\downarrow,\uparrow,\downarrow,\downarrow$''. On the right hand side a realistic interpretation of this experiment is sketched: despite of the apparent identical preparation procedures, at each of them the system randomly acquires the value $\uparrow$ or $\downarrow$ for the spin $s_z$ property which remains unknown till the measurement is performed. Measurements just act to reveal these preexistent values. Thus, realism asserts that when the quantum system is prepared in the spin state $|\psi\rangle$, what actually happens is that spin $s_z$ property is prepared in $\uparrow$ with probability $P(\uparrow)=|\tensor*[_z]{\langle \uparrow|\psi\rangle}{}|^2$ and in $\downarrow$ with probability $P(\downarrow)=|\tensor*[_z]{\langle \downarrow|\psi\rangle}{}|^2$. This would explain why these are the probabilities obtained for the measurement outcomes.}
\label{fig1}
\end{figure}

We may consider different hypotheses regarding the nature of quantum measurement outcomes. Probably, the most fundamental one is the assumption of \emph{realism}. In a broad sense, we can consider realism as \emph{the belief that there exists a physical reality independent of observers and measurement processes}. More specifically, \emph{the properties of physical systems have predefined values before its measurement, or even in the absence of it; measurements just act to reveal these values}. We illustrate this concept in Fig.~\ref{fig1}.

Realism is sometimes used as a synonym of \emph{determinism}, in the sense the physical observables have indeed \emph{determined} preexistent values. However this kind of determinism should not be confused with the assumption of \emph{causal determinism}, which states that the future values of the physical properties are completely determined by its present state. This happens for instance in classical mechanics. Here we will not assume that, we can consider quantum mechanics as an ultimate theory regarding predictability and still ask whether observables have predefined values before its measurement.

Besides realism, Bell inequalities and KS theorems consider local and noncontextual models, respectively. Specifically, the belief of
\emph{noncontextual realism} asserts that \emph{the predefined value of a real property is independent of the way the observer employs to measure it}. In particular, in the framework of the quantum mechanics, it implies that the (supposedly predefined) value of a quantum observable is independent of which other compatible observables are measured along with it. On the other hand, in the case of \emph{local realism}, the statement is that the \emph{value of an observable measured at some region of the Spacetime does not depend on whether or not another observable is measured in another region causally disconnected (i.e. space-like separated) from the first one}. In practical terms locality is a special case of noncontextuality, as observables from different regions of the Spacetime are compatible.

Finally the so-called \emph{measurement independence} or \emph{free will} assumption essentially says that \emph{the experimenter has the freedom to chose the measurement of whatever observable without affecting the supposed predefined values for the real properties of the system}, or viceversa \emph{the predefined values for the real properties does not condition the experimenter's measurement choice}. In other words, the values of the real properties are uncorrelated with the observer's choice of measurement settings \cite{Bellbook,Michael1}.

\section{Hidden Variable models and Joint Distributions}

Assuming that observable values correspond to real properties, under the standard setting of a hidden variable model \cite{Bellbook}, it is common to assume the existence of an (or several) underlying variable $\lambda$ which define the observed statistics of any observable $X$ by the equation
\begin{equation}\label{lambda1}
P_X(x)=\int\chi_X(x,\lambda)\rho(\lambda)d\lambda.
\end{equation}
Here $\chi_X(x,\lambda)$ is the indicator function for the real value $x$ for the observable $X$ given the value $\lambda$ for the underlying or ``hidden'' variable, and $\rho(\lambda)$ is a probability density function for its values. We shall consider the \emph{deterministic} case, otherwise $\chi_X(x,\lambda)$ must be replaced by a conditional probability. However, it has been shown by Fine \cite{Fine2}, and generalised in Appendix A of \cite{Michael1}, that there is no loss of generality in doing so. 

Given any other observable $Y$, the assumption of the existence of $\lambda$ and $\rho(\lambda)$ allows us to formulate a joint distribution for the observables $X$ and $Y$ in the form
\begin{equation}\label{lambda2}
P_{XY}(x,y)=\int\chi_X(x,\lambda)\chi_Y(y,\lambda)\rho(\lambda)d\lambda.
\end{equation}
This makes perfect physical sense if $X$ and $Y$ represent jointly measurable properties. However if $X$ and $Y$ corresponds to noncommuting quantum observables, it is natural to question whether such a model is possible to describe quantum experiments, as the definition of joint probability for incompatible quantum observables is known to present several problems \cite{Wigner,Dirac,Margenau,Cohen,Park}. Therefore, one may be not very surprised if a prediction based on this model, such as the original CHSH inequality \cite{CHSHpaper}, is violated by quantum mechanics. However, unless models must necessarily be of this form under the assumptions of realism, locality and measurement independence, nothing definitive can be concluded about whether these assumptions are ruled out by such a CHSH violation. It may instead be that some implicit additional assumption does not hold in nature. 

In this regard, the three main results of this work are: \\

\noindent i) There are alternative models for realism which avoid the existence of joint probabilities for incompatible quantum observables.\\

\noindent ii) Despite the absence of joint incompatible probabilities, it is possible to formulate a kind of CHSH inequality even for these models which is violated by quantum mechanics.\\

\noindent iii) There exist some specific limiting situations where it is impossible to assume noncontextual realism without assuming the existence of joint probabilities for incompatible quantum observables. \\

These points are developed in detail in the following sections.

\section{Realism without Incompatible Joint Distributions}
A realist model not allowing for joint distributions of incompatible observables can be better formulated from a frequentist point of view. Consider again $X$ and $Y$ to be incompatible quantum observables, and a set of $N$ quantum systems identically prepared. Under the hypothesis of realism, we can define $N(x,y)$ as the number of systems with value $x$ for the real property $X$ and value $y$ for the real property $Y$, prior to any measurement act. In other words, $N(x,y)$ is the number of systems such that \emph{if we chose} to perform a measurement of $X$, we obtain the value $x$ and, \emph{if we chose} the measurement of $Y$, we obtain the value $y$. Thus, trivially $N(x)=\sum_y N(x,y)$ and $N(y)=\sum_x N(x,y)$ are the number of systems with predefined value $x$ for the measurement of $X$, and $y$ for the measurement of $Y$, respectively. Note that we have explicitly avoided any assumption of whether or not $X$ and $Y$ are jointly measurable or compatible observables; $N(x,y)$ is the number of systems with two specific properties under \emph{individual} measurements. According to the experience, for $N$ large we must have stochastic convergence
\begin{equation}\label{marginal}
\frac{N(x)}{N}\sim P_X(x)\quad\text{and}\quad \frac{N(y)}{N}\sim P_Y(y), \quad N\gg1, 
\end{equation}
where $P_X(x)$ and $P_Y(y)$ are the predicted quantum probabilities for the observables $X$ and $Y$, respectively. Here, we have tacitly assumed measurement independence (or free will), as $N(x,y)$ is independent of which measurement setting will be eventually implemented by the observer.

Despite the fulfilling of \eqref{marginal}, one can construct models for $N(x,y)$ such that $N(x,y)/N$ does not converge to any definite function for large $N$.  To this end, consider first a particular example where $X$ and $Y$ are dichotomic observables $x=\pm1$ and $y=\pm1$, with mean values $\langle X\rangle$ and $\langle Y\rangle$, respectively; and consider two well-defined probability distributions $P^{A}(x,y)$ and $P^{B}(x,y)$ written in the form of
\begin{equation}
P^{A,B}(x,y)=\frac{1}{4}(1+\langle X\rangle x+\langle Y\rangle y+ C^{A,B}xy),
\end{equation}
with $C^{A,B}$ some numbers (given by the second order moments), such that both $P^{A}(x,y)$ and $P^{B}(x,y)$ have the same left and right marginals,
\begin{align}
\sum_{y}P^{A,B}(x,y)=\frac{1}{2}(1+\langle X\rangle x)=P_X(x),\\
\sum_{x}P^{A,B}(x,y)=\frac{1}{2}(1+\langle Y\rangle y)=P_Y(y).
\end{align}
We shall take $C^{A}$ to be different from $C^{B}$. For the sake of illustration, let us assume $\langle X\rangle=1/4$ and $\langle Y \rangle = 1/8$, and take
\begin{align}
&C^{A}=\langle X\rangle+\langle Y\rangle-1=-\tfrac{5}{8},\\
&C^{B}=1-\langle X\rangle+\langle Y\rangle=\tfrac{7}{8}.
\end{align}
These numbers are chosen to be the minimum and maximum value, respectively, compatible with the positivity of both $P^{A,B}(x,y)$. Now, consider a partition of the natural numbers into disjoint intervals $\mathds{N}=I_0\cup I_1\cup I_2\cup\ldots$, with 
\begin{align*}
I_0&=\{1\},\\
I_1&=\{2,3\},\\
I_2&=\{4,5,6,7\},\\
I_3&=\{8,9,10,11,12,13,14,15\},\\
\vdots&\\
I_k&=\{2^k,\ldots,2^{k+1}-1\}
\end{align*}
and define two sets:
\begin{align}\label{AB}
A=I_1\cup I_3\cup I_5\cup\ldots \text{ and }  B=I_0\cup I_2\cup I_4\cup\ldots,
\end{align}
so that $\mathds{N}= A\cup B$. Then, suppose an ensemble of varying size $N$, such that for $N\in A$, $(x,y)$ is a pair of random variables which samples the probability $P^A(x,y)$, and, for $N\in B$, $(x,y)$ samples the probability $P^B(x,y)$. Thus, as $N$ is increasing the tendency of the relative frequency $N(x,y)/N$ varies, depending on whether $N\in A$ or $N\in B$. As we will see, this induces an eternal oscillation in the relative frequency which is not approaching any number. However, the marginal relative frequencies $N(x)/N$ and $N(y)/N$ do approach the correct probabilities as both $P^{A,B}(x,y)$ have the same marginals. 

\begin{figure}[t]
\begin{center}
\includegraphics[width=1\columnwidth]{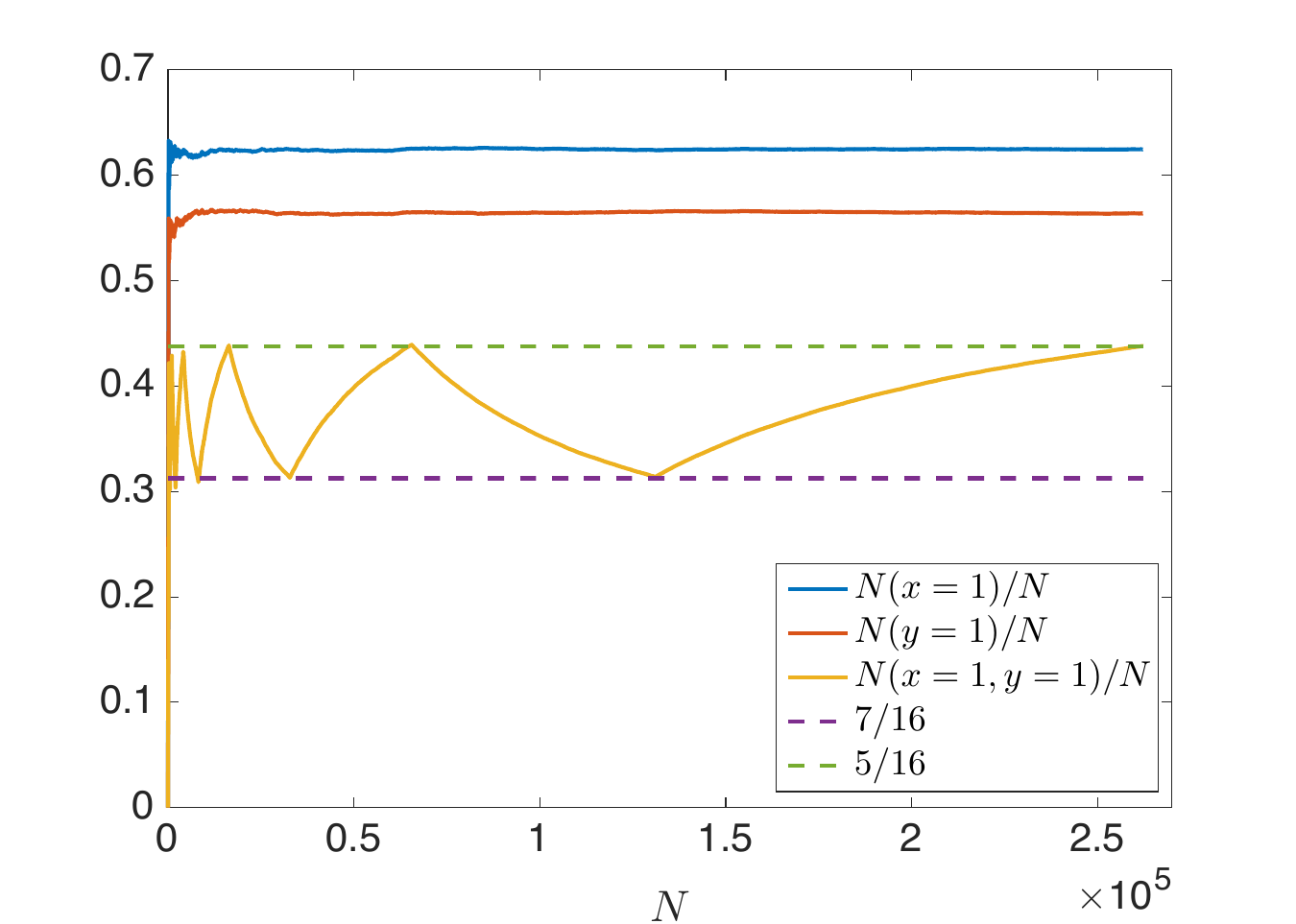}
\end{center}
\caption{Behavior of the relative frequencies $\frac{N(x=1)}{N}$, $\frac{N(y=1)}{N}$, and $\frac{N(x=1,y=1)}{N}$ as a function the ensemble size $N$ for the realist model described in the main text. Observables relative frequencies such as $\frac{N(x=1)}{N}$ and $\frac{N(y=1)}{N}$ approach the probabilities to find a system with real values $x=1$ (0.625) and $y=1$ (0.5625), respectively. However, for incompatible observables,  the nonobservable realist relative frequency $\frac{N(x=1,y=1)}{N}$ presents oscillations between $7/16$ and $5/16$ for large $N$. See discussion in main text.}
\label{fig2}
\end{figure}

We have plotted in Fig.~\ref{fig2} the relative frequencies $\tfrac{N(x=1)}{N}$, $\tfrac{N(y=1)}{N}$ and $\tfrac{N(x=1,y=1)}{N}$ for this model as a function of $N$. The first two show standard stochastic convergence, but the latter one oscillates between 
$\tfrac{7}{16}$ and $\tfrac{5}{16}$ for $N$ large enough.

In this example, the size of each interval $I_k$ is $2^k$, i.e. there are $2^k$ natural numbers inside the interval $I_k$. In general terms, and not restricting to dichotomic variables, we can consider intervals of size $m^k$ for some natural number $m$. We show in the Appendix that if the intervals have this exponential growth, then $N(x,y)/N$ oscillates between 
\begin{equation}
\frac{P^{A}(x,y)+mP^{B}(x,y)}{m+1}\text{ and }\frac{mP^{A}(x,y)+P^{B}(x,y)}{m+1}
\end{equation} 
for $N$ large. This corresponds to $\tfrac{7}{16}$ and $\tfrac{5}{16}$ in the above example for $m=2$. 

For this oscillatory behavior the exponential growth of the intervals becomes crucial. Convergence is obtained for intervals with a polynomial growing size (see Appendix). Therefore, in this kind of realist models with intervals with exponential growth, the joint distribution on $X$ and $Y$ is ill-defined despite the observable frequencies $N(x)/N$ and $N(y)/N$ converge to the correct probabilities $P_X(x)$ and $P_Y(y)$, respectively.

It is interesting to notice that in this model the number of systems $N(x,y)$ with the properties $(x,y)$ follows a sequence which is not Ces\`aro summable in the limit of $N\to\infty$, despite the sequences for $N(x)$ and $N(y)$ are Ces\`aro summable \cite{HardyBook}. Equivalently, from the perspective of the probabilistic number theory, the natural density of the sets $A$ and $B$ does not exist \cite{Tenenbaum}.

In summary, realism by itself does always not imply the existence of well-defined joint probabilities for arbitrary (including incompatible) pairs of observables.

\section{A CHSH inequality for finite ensembles}
The examples of the previous section show that standard derivations of Bell inequalities, which imply the existence of a joint probability distribution for sets of incompatible observables, such as in Eq. \eqref{lambda2}, thus have a potential loophole. Here this loophole is addressed by obtaining a CHSH-type inequality valid for those realist models.

Consider the standard scenario of a CHSH test, with $X$ and $X'$ for Alice and $Y$ and $Y'$ for Bob to be dichotomic observables with values $\pm1$. Thus, the $X$-observables are compatible with the $Y$-observables, but $X$ can be incompatible with $X'$, and $Y$ with $Y'$. For an ensemble of size $N$, the quantity $N(x,x',y,y')$, defined as before for individual measurements results, and its marginals exist under the hypothesis of realism. Similarly as before, we can
consider
\begin{align}
P^{A,B}(x,x',y,y')=\frac{1}{16}\left(1+\sum_i\langle O_i\rangle o_i+\sum_{i<j}C^{A,B}_{ij}o_io_j\right.\nonumber\\
\left.+\sum_{i<j<k}C^{A,B}_{ijk}o_io_jo_k+C^{A,B}_{1234}o_1 o_2 o_3 o_4\right)
\end{align}
with the identification $O_1=X$, $O_2=X'$, $O_3=Y$, $O_4=Y'$ and $o_1=x$, $o_2=x'$, $o_3=y$, $o_4=y'$. Here $C_{ij}^{A,B}$, $C_{ijk}^{A,B}$ and $C_{1234}^{A,B}$ are numbers (given by the second, third and fourth order moments, respectively). If we construct the sets $A$ and $B$ as in \eqref{AB} with intervals $I_k$ of exponential growth, such that 
\begin{equation*}\begin{array}{ll}
C_{ij}^{A}=C_{ij}^{B}, & \text{iff $O_i$ and $O_j$ compatible,}  \\[\medskipamount]
C_{ijk}^{A}=C_{ijk}^{B},  &\text{iff $O_i, O_j$ and $O_k$ are compatible,} \\[\medskipamount]
C_{1234}^{A}=C_{1234}^{B}, & \text{iff $O_1, O_2, O_3$ and $O_4$ are compatible,}
\end{array}
\end{equation*}
any joint relative frequency of compatible observables has a limit, and any joint relative frequency involving a pair of incompatible observable does not. In particular, if some pair of noncommuting observables are considered in the set $\{X,X',Y,Y'\}$, then $N(x,x',y,y')/N$ is not convergent and so a formulation in terms of equations like \eqref{lambda1} and \eqref{lambda2} is problematic. Note that, in such a situation, $\rho(\lambda)$ in \eqref{lambda1} and \eqref{lambda2} may not exist. This is clear if $\lambda$ just label the different instances of predefined real values, $\lambda=(x,x',y,y')$, as $\rho(\lambda)$ would represent the joint probability function.

One may wonder whether Bell inequalities can be obtained for a model with ill-defined joint distributions, or whether such a model can provide an instance of a kind of realism compatible with quantum mechanics. The answer to this last statement is negative. 
To obtain a Bell inequality for models of this type, consider the following equation
\begin{multline}
\sum_{x,y,x',y'=\pm1}[(x+x')y+(x-x')y']\frac{N(x,x',y,y')}{N} \\
=\langle XY\rangle_N+\langle XY'\rangle_N+\langle X'Y\rangle_N-\langle X'Y'\rangle_N.
\end{multline}
where
\begin{align}
&\langle XY\rangle_N:=\sum_{x,y,x',y'}xy\tfrac{N(x,x',y,y')}{N}=\sum_{x,y}xy\tfrac{N(x,y)}{N},\label{EqXY}\\
&\langle XY'\rangle_N:=\sum_{x,y,x',y'}xy'\tfrac{N(x,x',y,y')}{N}=\sum_{x,y'}xy'\tfrac{N(x,y')}{N},\\
&\langle X'Y\rangle_N:=\sum_{x,y,x',y'}x'y\tfrac{N(x,x',y,y')}{N}=\sum_{x',y}x'y\tfrac{N(x',y)}{N},\\
&\langle X'Y'\rangle_N:=\sum_{x,y,x',y'}x'y'\tfrac{N(x,x',y,y')}{N}=\sum_{x',y'}x'y'\tfrac{N(x',y')}{N}\label{EqX'Y'},
\end{align}
are the values of the correlations for an ensemble of size $N$. Then, following a standard argument, $|(x+x')y+(x-x')y'|=2$ for $x,y,x',y'\in\{-1,1\}$, and $\frac{N(x,x',y,y')}{N}$ is positive and sums to $1$ for any $N$, so we obtain
\begin{equation}\label{CHSHfinite}
-2\leq\langle XY\rangle_N+\langle XY'\rangle_N+\langle X'Y\rangle_N-\langle X'Y'\rangle_N\leq2.
\end{equation}
This looks like a finite size version of the CHSH inequality but one should note that it is actually purely formal. No assumptions about \emph{locality} or \emph{measurement independence} have been considered so far. Inequality \eqref{CHSHfinite} holds for the mean values computed by Eqs. \eqref{EqXY}-\eqref{EqX'Y'} for any ensemble of $N$ systems. However, in order to obtain an inequality which can be compared with physical predictions, those hypothesis are needed.

Specifically, under locality, the Alice's action of measuring some observable, say $X$, finding some outcome $x$, does not change the predefined value $y$ for a Bob's measurement of $Y$ on that system. In the same way, the Bob action of measuring $Y$, finding some outcome $y$, does not modify the predefined value $x$ for the Alice's observable $X$ on that system. Thus, the assumption of locality allows Alice and Bob to make a \emph{simultanous} measurement of $X$ and $Y$, to determine the number $N(x,y)$ of systems in the ensemble with predefined values $x$ and $y$ for \emph{individual} measurements of $X$ and $Y$, respectively. The same hold for the pairs $X$ and $Y'$, $X'$ and $Y$, and $X'$ and $Y'$. In addition, under measurement independence, the number $N(x,x',y,y')$ is \emph{intrinsic} to the system and the preparation procedure, and \emph{independent} of what posterior measurement scheme is implemented in the Alice and Bob's laboratories. In fact, we may allow $N(x,x',y,y')$ to be measurement dependent provided that the four numbers $N(x,y)$, $N(x,y')$, $N(x',y)$, and $N(x',y')$ \emph{are not}. In order words, if we specify by $N_M(x,y,x',y')$ the number systems with predefined values for $\{X,X',Y,Y'\}$ under the measurement choice $M$; measurement independence asserts that $N_M(x,y,x',y')=N(x,y,x',y')$, but according to Eqs. \eqref{EqXY}-\eqref{EqX'Y'}, only
\begin{align}
N(x,y)=N_M(x,y)=\sum_{x',y'}N_M(x,y,x',y'),\\
N(x',y)=N_M(x',y)=\sum_{x,y'}N_M(x,y,x',y'),\\
N(x,y')=N_M(x,y')=\sum_{x',y}N_M(x,y,x',y'),\\
N(x',y')=N_M(x',y')=\sum_{x,y}N_M(x,y,x',y'),
\end{align}
are required. Then we can drop any subindex on the choice of measurement setting $M$ (as we tacitly did) in Eq. \eqref{CHSHfinite}.

The measurement independence assumption is needed to formulate a test of \eqref{CHSHfinite} because Alice and Bob cannot make a simultaneous measurement of the four observables $X$, $X'$, $Y$ and $Y'$; as $X$ and $X'$ or $Y$ and $Y'$ may be incompatible. Thus, in an hypothetical situation, Alice and Bob may choose to make a measurement $M_1$ of the pair $X$ and $Y$, then for large enough $N$, the observed $\langle XY\rangle_N$ must approach the quantum prediction. However, they could have chosen a measurement $M_2$ of the pair $X$ and $Y'$ instead, and then in such a case they would have found that $\langle XY'\rangle_N$ approaches the quantum prediction for large $N$. The same applies for $\langle X'Y\rangle_N$ and $\langle X'Y'\rangle_N$, depending on what measurement scheme $M$ Alice and Bob decide to perform. Since measurement independence asserts that $N(x,y)$, $N(x,y')$, $N(x',y)$, and $N(x',y')$ does not depend on $M$, provided that $N$ is large enough, quantum predictions for the four mean values can be introduced in the inequality \eqref{CHSHfinite}, forgetting any choice of the measurement scheme. 

All this allows us to check the bound \eqref{CHSHfinite} by using the quantum predition for a simulatenous measurement of pairs of spin components $X=\sigma_{\bm{a}}\otimes \mathds{1}$, $X'=\sigma_{\bm{a}'}\otimes \mathds{1}$, $Y=\mathds{1}\otimes \sigma_{\bm{b}}$ and $Y=\mathds{1}\otimes \sigma_{\bm{b}'}$, in a bipartite system prepared in a singlet state. This leads to the the well-known violation for large $N$ \cite{Bellbook,CHSHpaper}. Hence, this closes the potential loophole to the Bell theorem.

It is worth to mention that, strictly speaking, the argument above formulates a theoretical violation, based on an hypothetical case, or gedankenexperiment. An experimental confirmation of the violation of the inequality \eqref{CHSHfinite} needs however an extra further assumption. If Alice and Bob want to measure the four mean values, they need to split the original ensemble of size $N$ into four subensembles of smaller size, each one intended to be subject to a simultaneous measurement of an observable pair. The size of these subensembles must increase linearly with $N$, so we can write $N_j=f_j N$ for the size of the subensemble $j$, with $1>f_j>0$. The required assumption is that \emph{for any pair of compatible observables, say $X$ and $Y$, if $N(x,y)$ is the number of systems with predefined values $x$ and $y$ in the ensemble of size $N$, then for large enough $N$, the number of systems $N_j(x,y)$ with predefined values $x$ and $y$ in the subensemble of size $N_j=f_j N$ must satisfy $N_j(x,y)\approx f_j N(x,y)$}. This assumption ensures that
\begin{equation}
\frac{N(x,y)}{N}\approx \frac{N_j(x,y)}{N_j}, \quad\text{for } N\gg1,
\end{equation}
and the same for $N(x,y')$, $N(x',y)$, and $N(x',y')$, so that
\begin{align*}
&\langle XY\rangle_N\approx\sum_{x,y}xy\tfrac{N_j(x,y)}{N_j},\\
&\langle XY'\rangle_N\approx\sum_{x,y'}xy'\tfrac{N_j(x,y')}{N_j},\\
&\langle X'Y\rangle_N\approx\sum_{x',y}x'y\tfrac{N_j(x',y)}{N_j},\\
&\langle X'Y'\rangle_N\approx\sum_{x',y'}x'y'\tfrac{N_j(x',y')}{N_j},
\end{align*}
for large enough $N$. Therefore these numbers become observables quantities dividing the original ensemble of $N$ systems into four smaller subensembles. 

Nevertheless, this assumption is apparently rather weak because, in a quantum measurement on a sample of $N\gg 1$ systems, the relative frequency of results is approximately the same in any subset which is constructed by taking systems randomly. Moreover, it is usually supposed that the observer has the ability to select systems randomly, at least to some extent. However, note this reasoning does not work for nonobservable objects involving incompatible observables, such as $N(x,x',y,y')$. Actually, if $N(x,x',y,y')/N$ is nonconvergent we should not expect an equation like $N_j(x,x',y,y')\approx f_jN(x,x',y,y')$ to be true, regardless of whether or not measurement independence holds. 

To summarize, despite there are realist models which prevents from the existence of well-defined joint probabilities for noncommuting observables, they are not compatible with quantum mechanics under assumptions based on locality and measurement independence because they actually satisfy a finite ensemble version of the CHSH inequality \eqref{CHSHfinite}. 

\section{Joint distribution for incompatible observables imposed by noncontextual realism}
It has been shown that standard derivations of Bell theorems such as the CHSH inequality have a loophole concerning the existence of well-defined joint probability distributions for incompatible observables, but that this loophole can be closed by deriving such theorems for strictly finite ensembles.  In contrast, we show here that certain Kochen-Specker theorems do not admit, in the first place, realists models with ill-defined joint probability distribution for any set of incompatible observables. 

A case of this can be seen in a Hardy-like proof of quantum contextuality in \cite{CabelloHardy} which we shall adapt to our case. Essentially, in this scheme we have an ensemble of $N$ three level systems prepared in the pure state vector $|\psi\rangle=\frac{1}{\sqrt{3}}(1,1,1)^{\rm T}$. One considers the five projector observables $\{\Pi_j=|v_j\rangle\langle v_j|\}_{j=1}^5$ with
\begin{align*}
&\ket{v_1}=\frac{1}{\sqrt{3}}(1,-1,1)^{\rm T},\\
&\ket{v_2}=\frac{1}{\sqrt{2}}(1,1,0)^{\rm T}, \\
&\ket{v_3}=(0,0,1)^{\rm T},\\
&\ket{v_4}=(1,0,0)^{\rm T}, \\
&\ket{v_5}=\frac{1}{\sqrt{2}}(0,1,1)^{\rm T}.
\end{align*}
It is straightforward to check that $\{\Pi_1,\Pi_2\}$, $\{\Pi_2,\Pi_3\}$, $\{\Pi_3,\Pi_4\}$, $\{\Pi_4,\Pi_5\}$, and $\{\Pi_1,\Pi_5\}$ are sets of compatible obsevables. Moreover the probability to obtain the value 1 for both $\Pi_1$ and $\Pi_2$ is 
\begin{equation}
P_{1,2}(1,1)=\bra{\psi}\Pi_1\Pi_2\ket{\psi}=0,
\end{equation}
and similarly $P_{2,3}(1,1)=P_{2,3}(0,0)=P_{3,4}(1,1)=P_{4,5}(1,1)=P_{4,5}(0,0)=P_{1,5}(1,1)=0$. Under the hypotheses of realism, and following the same notation as previous sections, we can write $N_{1,2}(1,1)$ for the number of systems with predefined value 1 for both real properties associated to the (individual) measurement of $\Pi_1$ or $\Pi_2$, in the ensemble of size $N$. So that we have
\begin{multline}\label{Hardy0}
N_{1,2}(1,1)=N_{2,3}(1,1)=N_{2,3}(0,0)=N_{3,4}(1,1)\\
=N_{4,5}(1,1)=N_{4,5}(0,0)=N_{1,5}(1,1)=0,
\end{multline}
for any size $N$. However, if the predefined value of an observable depends neither on which other (if any) compatible observable is measured along with it (noncontextual realism), nor on the observer's choice on which pair is going to be measured (measurement independence), then 
\begin{equation}\label{Hardy01}
N_{1,5}(1,0)=0.
\end{equation}
Indeed, suppose that a system in the ensemble has predefined value 1 for $\Pi_1$, then it has 0 for $\Pi_2$ [as $N_{1,2}(1,1)=0$], then it has 1 for $\Pi_3$ [as $N_{2,3}(0,0)=0$], then it has 0 for $\Pi_4$ [as $N_{3,4}(1,1)=0$], and then has 1 for $\Pi_5$ [as $N_{4,5}(0,0)=0$], therefore $N_{1,5}(1,0)=0$. However for large enough $N$, we have
\begin{equation}\label{N15}
\frac{N_{1,5}(1,0)}{N}\sim P_{1,5}(1,0)=\bra{\psi}\Pi_1(\mathds{1}-\Pi_2)\ket{\psi}=\frac{1}{9}\neq0,
\end{equation}
and this is the desired contradiction.

To see that this ensemble under noncontextual realism imposes the existence of a well-defined joint distribution for incompatible observables, let us focus in a pair such as $\Pi_1$ and $\Pi_3$, $[\Pi_1,\Pi_3]\neq0$. Paraphrasing the previous argument, under noncontextual realism and measurement independence we conclude that
\begin{equation}\label{N13}
N_{1,3}(1,0)=0.
\end{equation}
This implies that $N_{1,3}(i,j)/N$ converges with probability 1 for large $N$ to some well-defined probability function $P_{1,3}(i,j)$:
\begin{equation}
\lim_{N\to\infty}\frac{N_{1,3}(i,j)}{N}=P_{1,3}(i,j).
\end{equation}
This can be easily shown. First of all, $P_{1,3}(1,0)=0$ trivially. Secondly, since $N_{1,3}(0,0)+N_{1,3}(1,0)=N_{3}(0)$ which is an observable quantity,
\begin{equation}
P_{1,3}(0,0)=\lim_{N\to\infty}\frac{N_{1,3}(0,0)}{N}= \lim_{N\to\infty}\frac{N_{3}(0)}{N}=P_3(0).
\end{equation}
Similarly,
\begin{equation}
P_{1,3}(1,1)=\lim_{N\to\infty}\frac{N_{1,3}(1,1)}{N}= \lim_{N\to\infty}\frac{N_{1}(1)}{N}=P_1(1),
\end{equation}
and finally by using $N_{1,3}(0,1)=N-N_{1,3}(0,0)-N_{1,3}(1,0)-N_{1,1}(0,1)$,
\begin{align}
P_{1,3}(0,1)&=\lim_{N\to\infty}\frac{N_{1,3}(0,1)}{N}\nonumber \\
&=1-P_{1,3}(0,0)-P_{1,3}(1,0)-P_{1,1}(0,1).
\end{align}
Hence, in this case, noncontextual realism imposes the existence of $P_{1,3}(i,j)$ to be a well-defined joint probability function for the quantum observables $\Pi_1$ and $\Pi_3$ despite they do not commute and the state vector $\ket{\psi}$ is not an eigenvector of either $\Pi_1$ or $\Pi_3$. This makes a difference with the previous example in the CHSH setting where it is possible to find models where $N(x,x')/N$ and $N(y,y')/N$ does not approach to any number for incompatible $X$ and $X'$, and $Y$ and $Y'$, respectively.

However, one should notice that for this set of observables $\{\Pi_j\}_{j=1}^{5}$, the only state fulfilling $P_{2,3}(0,0)=P_{4,5}(0,0)=0$ is the one considered here $\ket{\psi}=\tfrac{1}{\sqrt{3}}(1,1,1)^{\rm T}$ (up to some global phase factor). So that, the conditions leading to the equalities \eqref{Hardy01} and \eqref{N13} for any system size can be considered as a limiting case of a zero-measure set. This of course does not impose the existence of joint probabilities in practice for the preparations and observables subject to finite precision.

\section{Discussion}
As seen, realism by itself does not assume the existence of joint probabilities for noncommuting observables, as realist models can be formulated without these objects. However, Bell inequalities can be derived for finite size ensembles under assumptions based on local realism and measurement independence. They rule out those models as valid realist descriptions of quantum mechanics, even if they share with the quantum theory the property of lacking joint probabilities for incompatible observables. 

In the considered model, relative frequencies oscillate for incompatible observables, but converge for compatible (or individual) ones. This nonconvergence of relative frequencies for two real properties might be seen as a kind of bizarre property. This feeling can be enforced by the mathematical fact that a sequence of numbers without a convergent relative frequency, as in our model, is an instance of divergent series that is not Ces\`aro summable \cite{HardyBook}. Nevertheless, nothing prevents nature to behave that way. Nonconvergent sequences are present in some physical theories, such as quantum field theory, and the one considered here is not linked to any directly observable object. The fact that the finite ensemble CHSH inequality can also rule out such a realist model, stresses the importance of the Bell theorem and the role of local realism and measurement independence.

In this regard, we must point how these results connect with the Fine's theorem \cite{Fine1,Fine2}. It asserts that four observables $\{X,X',Y,Y'\}$ which fulfill the CHSH inequality (with all possible permutations between $X$ and $X'$, and between $Y$ and $Y'$) admits a valid joint probability distribution. Our model present an ill-defined joint probability distribution but satisfy the inequality \eqref{CHSHfinite} for any $N$. Then, Fine's theorem just implies that a well-behaved joint probability $P(x,y,x',y')$ can be constructed with the same marginals as our model for the observable probabilities, e.g. $P(x,y)=\lim_{N\to\infty}N(x,y)$. However any nonobservable probability constructed from $P(x,y,x',y')$ is purely formal, and need not reflect the actual physical situation described by the realist model. Actually, one could also construct $P(x,y,x',y')$ for a classical correlation scenario, but the true classical situation will be almost certainly different from the Fine's model.

On the other hand, it is interesting to note that, since we may permit $N(x,x',y,y')$ to be measurement dependent provided that the four numbers $N(x,y)$, $N(x,y')$, $N(x',y)$, and $N(x',y')$ are not, the finite ensemble CHSH inequality is still valid even under a weak absence of free will which conspiratorially affects just to nonobservable relative frequencies. Conditions under which some degree of measurement dependence (on observable objects) can be allowed not spoiling the Bell theorem has been studied in \cite{Michael1}.

Finite sampling arguments to rule out realism appeared in the literature time ago \cite{Stapp}. Mermin \cite{Mermin2} and Macdonald \cite{Macdonald} employed some inequalities based on relative frequencies to discard the Pitowsky's hidden variable model in \cite{Pitowsky}. In such a model, equations such as \eqref{lambda2} are formulated for compatible observables, e.g. a pair of spin components for Alice and Bob. However the equality \eqref{lambda1} fails because $\lambda$ takes values on a nonmeasurable set. Of course, this model is ruled out by our Eq. \eqref{CHSHfinite} as well. Also Gill derived a probabilistic version of the CHSH innequality for finite sampling \cite{Gill}. He based his derivation on observable averages and assuming, as we commented in Sec. V, that starting from an ensemble of $N$ systems the observer can construct four exclusive subsets randomly, with equally distributed relative frequencies. Nevertheless, to our knowledge, it has not be presented a detailed statement and analysis of a finite ensemble CHSH inequality \eqref{CHSHfinite} under the potential case of nonconvergence for incompatible relative frequencies, discussing the form that locality and measurement dependence take in this scenario.

Finally, we have given an instance where noncontextual realism impose the existence of joint distributions for incompatible observables. A connection between the existence of valid joint distributions and a kind of general noncontextual models was derived in \cite{Spekkens2}. One could wonder whether locality (instead of noncontextuality) may also imply the existence of joint distributions for incompatible observables in some specific quantum setting. The answer is affirmative. This can be easily checked, following the same steps as in Sec. VI, in the original formulation of the Hardy paradox where locality is explicitely assumed \cite{Hardy1}. However, in both cases, the argument turns out to be fragile as only works for a specific state which is impossible to prepare with unity fidelity in real experiments. For all these reasons, we believe that the existence of joint probabilities should be definitively considered as a different assumption than realism.

\section*{Acknowledgments}
The author is in debt with Michael J. W. Hall for many enlightening discussions and for pointing him a previous mistake in a preliminary version of this work. He also thanks Alfredo Luis for several helpful comments.
Partial financial support by the Spanish MINECO grants FIS2012-33152, FIS2015-67411, and FIS2017-91460-EXP, the CAM research consortium QUITEMAD+ grant S2013/ICE-2801, and the US Army Research Office through Grant No. W911NF-14-1-0103 is acknowledged.

\onecolumngrid
\section{Appendix}
We provide here further details about the oscillatory behavior of $N(x,y)/N$ in the model of the main text, and why intervals with an exponential growing size are required to avoid convergence.  

If $N_k:=m^k$ is the number of numbers in the set $I_k$, take, without loss of generality, $N$ to be the sum of an integer number intervals, $N=N_1+N_2+\ldots+N_k$, so that  
\begin{equation}
\frac{N(x,y)}{N}= \frac{N_1(x,y)+N_2(x,y)+\ldots+N_k(x,y)}{N_1+N_2+\ldots+N_k},
\end{equation}
with $N_k(x,y)$ the number of systems with predefined values $(x,y)$ in the set associated to the interval $I_k$. 
Then for $N$ large enough (i.e. $k$ large enough), we have 
\begin{equation}
N_k(x,y)\approx\begin{cases}
P^A(x,y)N_k, &\text{if $k$ is odd},\\
P^B(x,y)N_k, &\text{if $k$ is even}.
\end{cases}
\end{equation}
Thus, we obtain 
\begin{equation}
\frac{N(x,y)}{N}\approx\left\{\begin{array}{ll}
\frac{P^A(x,y) N_1+ P^B(x,y)N_2+\ldots+P^A(x,y) N_k}{N_1+N_2+\ldots+N_k}, & \text{if $k$ is odd},\\
\\
\frac{P^A(x,y) N_1+ P^B(x,y)N_2+\ldots+P^B(x,y) N_k}{N_1+N_2+\ldots+N_k}, & \text{if $k$ is even}.
\end{array}\right.
\end{equation}
The point is that both expressions approach to different numbers. To see this, on the one hand, provided that $m\neq1$ the sum of the geometric progression gives
\begin{equation}
N_1+N_2+\ldots+N_k=m^1+m^2+\ldots+m^k=\frac{m-m^{k+1}}{1-m},
\end{equation}
and similarly,
\begin{align}
&N_2+N_4+\ldots+N_{2n}=m^2+m^4+\ldots+m^{2n}=\frac{m^2-m^{2n+2}}{1-m^2},\\
&N_1+N_3+\ldots+N_{2n+1}=m^1+m^3+\ldots+m^{2n+1}=\frac{m-m^{2n+2}}{1-m}-\frac{m^2-m^{2n+2}}{1-m^2}=\frac{m-m^{2n+3}}{1-m^2},
\end{align}
Now, suppose $k$ to be even,
\begin{align}
P^A(x,y)N_1 +P^B(x,y)N_2+\ldots+P^B(x,y) N_k
&=P^A(x,y)(m+m^3+\ldots m^{k-1})+ P^B(x,y)(m^2+m^4+\ldots+m^k)\nonumber\\
&=P^A(x,y)\frac{m-m^{k+1}}{1-m^2}+P^B(x,y)\frac{m^2-m^{k+2}}{1-m^2},
\end{align} 
this yields
\begin{equation}\label{mPB}
\frac{N(x,y)}{N}\approx \frac{1-m}{m-m^{k+1}}\left[P^A(x,y)\frac{m-m^{k+1}}{1-m^2}+P^B(x,y)\frac{m^2-m^{k+2}}{1-m^2}\right]=\frac{P^A(x,y)+mP^B(x,y)}{m+1}.
\end{equation}
However, suppose $k$ to be odd
\begin{align}
P^A(x,y)N +P^B(x,y)N_2+\ldots+P^A(x,y) N_k
&=P^A(x,y)(m+m^3+\ldots m^k)+ P^B(x,y)(m^2+m^4+\ldots+m^{k-1})\nonumber\\
&=P^A(x,y)\frac{m-m^{k+2}}{1-m^2}+P^B(x,y)\frac{m^2-m^{k+1}}{1-m^2},
\end{align} 
and in this case the relative frequency becomes 
\begin{align}\label{mPA}
\frac{N(x,y)}{N}&\approx \frac{1-m}{m-m^{k+1}}\left[P^A(x,y)\frac{m-m^{k+2}}{1-m^2}+P^B(x,y)\frac{m^2-m^{k+1}}{1-m^2}\right]\nonumber \\
&=\frac{P^A(x,y)}{m+1}\left[\frac{1-m^{k+1}}{1-m^k}\right]+\frac{P^B(x,y)}{m+1}\left[\frac{m-m^{k}}{1-m^k}\right]\xrightarrow{k\gg1}\frac{mP^A(x,y)+P^B(x,y)}{m+1}.
\end{align}
Therefore for large $N$, the relative frequencies oscillate between both quantities in Eqs. \eqref{mPB} and \eqref{mPA} as claimed.

It is interesting to see that the size of the intervals $I_k$ must grows exponentially to avoid convergence. For instance, suppose instead that the size of $I_k$ grows like a power $k^\alpha$, $\alpha\in\mathds{N}$. Then for the sum of powers of natural numbers we have \cite{Tables}
\begin{equation}
1^\alpha+2^\alpha+\ldots+k^\alpha=\frac{k^{\alpha+1}}{\alpha+1}+\frac{k^\alpha}{2}+\frac{1}{2}\binom{\alpha}{1}B_2 k^{\alpha-1}+\frac{1}{4}\binom{\alpha}{3}B_4 k^{\alpha-3}+\frac{1}{6}\binom{\alpha}{5}B_6 k^{\alpha-5}+\ldots, 
\end{equation}
where $B_k$ are Bernouilli numbers. So, in this case for $k$ even
\begin{align}
P^A(x,y)N_1 +P^B(x,y)N_2+\ldots+P^B(x,y) N_k
&=P^A(x,y)[1^\alpha+3^\alpha+\ldots +(k-1)^\alpha]+ P^B(x,y)(2^\alpha+4^\alpha+\ldots+k^\alpha)\nonumber\\
&=P^A(x,y)\left[\frac{k^{\alpha+1}}{\alpha+1}-\frac{2^\alpha k^{\alpha+1}}{(\alpha+1)2^{\alpha+1}}\right]+ P^B(x,y)\frac{2^\alpha k^{\alpha+1}}{(\alpha+1)2^{\alpha+1}}+\mathcal{O}(k^\alpha)\nonumber\\
&=\frac{P^A(x,y)+ P^B(x,y)}{2}\frac{k^{\alpha+1}}{\alpha+1}+\mathcal{O}(k^\alpha)
\end{align} 
and similarly for $k$ odd
\begin{align}
P^A(x,y)N_1 +P^B(x,y)N_2+\ldots+P^A(x,y) N_k
&=P^A(x,y)[1^\alpha+3^\alpha+\ldots k^\alpha]+ P^B(x,y)[2^\alpha+4^\alpha+\ldots+(k-1)^\alpha]\nonumber\\
&=P^A(x,y)\left[\frac{k^{\alpha+1}}{\alpha+1}-\frac{2^\alpha (k-1)^{\alpha+1}}{(\alpha+1)2^{\alpha+1}}\right]+ P^B(x,y)\frac{2^\alpha (k-1)^{\alpha+1}}{(\alpha+1)2^{\alpha+1}}+\mathcal{O}(k^\alpha)\nonumber\\
&=\frac{P^A(x,y)+ P^B(x,y)}{2}\frac{k^{\alpha+1}}{\alpha+1}+\mathcal{O}(k^\alpha).
\end{align} 
Therefore, for $N$ large and equal to the sum of an integer number of intervals
\begin{equation}\label{meanP}
\frac{N(x,y)}{N}= \frac{N_1(x,y)+N_2(x,y)+\ldots+N_k(x,y)}{N_1+N_2+\ldots+N_k}\approx \frac{\tfrac{P^A(x,y)+ P^B(x,y)}{2}\tfrac{k^{\alpha+1}}{\alpha+1}}{\tfrac{k^{\alpha+1}}{\alpha+1}}=\frac{P^A(x,y)+ P^B(x,y)}{2},
\end{equation}
for both $k$ odd or even. For any large enough $N=N_1+N_2+\ldots+qN_k$, $1\geq q\geq0$ the relative frequency must be in between the relative frequency for $N=N_1+N_2+\ldots+N_k$ and $N=N_1+N_2+\ldots+N_{k-1}$, which in both cases approach the arithmetic mean of both distributions \eqref{meanP}. Therefore the relative frequency approaches to the valid probability distribution $\tfrac{P^A(x,y)+ P^B(x,y)}{2}$. This result shows that the size of the partitions in the model is a highly nontrivial issue.

\vspace{1cm}

\twocolumngrid

\end{document}